# Solutions to the GSM Security Weaknesses [†]

Mohsen Toorani [‡]          Ali A. Beheshti

## Abstract

*Recently, the mobile industry has experienced an extreme increment in number of its users. The GSM network with the greatest worldwide number of users succumbs to several security vulnerabilities. Although some of its security problems are addressed in its upper generations, there are still many operators using 2G systems. This paper briefly presents the most important security flaws of the GSM network and its transport channels. It also provides some practical solutions to improve the security of currently available 2G systems.*

## 1. Introduction

The Mobile communications has experienced a great acceptance among the human societies. It has influenced and revolutionized different aspects of the human life. With a mobile handset, anyone can be accessed anywhere. At the beginning of 2007, the worldwide number of mobile users reached to 2.83 billion people where 2.28 billion users out of them (i.e. 80.5%) were using the *Global Service for Mobile communications* (GSM) [1]. The GSM system and its building blocks are depicted in Figure 1. The GSM has experienced gradual improvements that leaded to several versions such as GSM1800, HSCSD (*High Speed Circuit Switched Data*), EDGE (*Enhanced Data rates for GSM Evolution*), and GPRS (*General Packet Radio Service*). The GSM improvements are continued to 3G systems such as UMTS. It is believed that the GSM has many inherent security flaws and some of its security flaws are addressed in the upper generations such as UMTS. However, many operators especially in the developing countries are still using the traditional GSM network that succumbs to several security flaws.

Although the GSM security is considered in some literatures [2-4], they did not present a complete security evaluation or even propose solutions. This paper provides a brief and complete review of the GSM security flaws, and some applicable solutions to improve the security of currently available GSM systems. This paper is organized as follows. Section 2 briefly introduces the security attributes of the GSM. Section 3 challenges the GSM security. The security of GSM transport channels is briefly described in section 4. Section 5 provides some practical solutions to improve the security of GSM, and section 6 gives the conclusions.

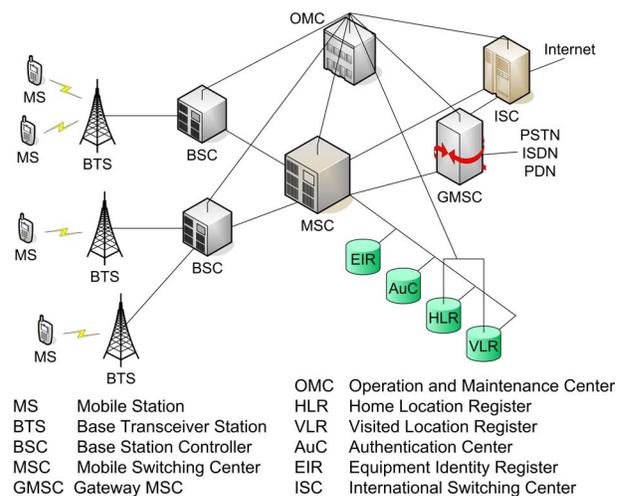

**Figure 1. GSM Architecture**

## 2. Security Architecture of the GSM

The security architecture of GSM was originally intended to provide security services such as anonymity, authentication, and confidentiality of user data and signaling information [5]. The security goals of GSM are as follows:

- Authentication of mobile users for the network,
- Confidentiality of user data and signaling information,
- Anonymity of subscriber's identity,
- Using SIM (*Subscriber Identity Module*) as a security module.

The *Mobile Station* (MS) consists of the *Mobile Equipment* (ME), and the SIM card. The SIM is a cryptographic smart card with the GSM specific applications loaded onto it. As a smart card, it has some inherent security functions specified to smart cards [6]. Its operating system and chip hardware have several security attributes. SIM includes all the necessary information to access the subscriber's account. IMSI and Ki are stored on every SIM. IMSI is the *International Mobile Subscriber Identity* with at most 15 digits uniquely devoted to every mobile subscriber in the world. Ki (Individual subscriber authentication Key) is a random 128-bits number that is the root cryptographic key used for generating session keys, and authenticating the mobile users to the network. Ki is strictly protected and is stored on the subscriber's SIM, and AuC. The SIM is itself protected by an optional *Personal Identification Number* (PIN). Each user is requested to enter the PIN unless this feature is deactivated by the user. After a number of invalid attempts that is usually 3 times, the SIM locks out the PIN, and the PUK (PIN UnlocK) is then requested. If the PUK is also incorrectly entered for a number of times that is usually 10 times, the SIM refuses local accesses to its privileged information and authentication functions, and makes itself useless.

Authentication and confidentiality of user data are in deposit of the secrecy of IMSI and Ki. With disclosure of such numbers, anyone can impersonate a legitimate user. A3 and A8 algorithms are also implemented on every SIM. This means that each operator can determine and change such algorithms independent of other operators and hardware manufacturers. Therefore, the authentication will work when a user is roaming on other countries or operators since the local network will query the HLR of the home network for the results and does not need to know the A3/A8 algorithm of the home network. A3 is mainly used for authenticating users to the network while A8 is used for generating the session key of encryption Kc. The network sends a random challenge to the user so that SIM produces Kc and SRES. After user authentication, the network can order the phone to start the encryption by using the generated session key Kc.

The cryptographic algorithms are implemented on the hardware of mobile phones. The network can choose from up to 7 different encryption algorithms (or the mode of no ciphering) but it should choose an algorithm that is implemented on the phones. A classmark message has been earlier specified the phone's capabilities to the network. Three algorithms are generally available: A5/1, A5/2, and A5/3. A5/1 and A5/2 are two stream ciphers originally defined by the GSM standards. A5/1 is stronger but it is subject to export control and can be used by those countries that are members of CEPT. A5/2 is deliberately weakened to be deployed by the other countries. The use of such algorithms is controlled by the GSM *Memorandum of Understanding* (MoU). A5/3 is a block cipher based on the Kasumi algorithm that is defined by the 3GPP at 2002 and can be supported on dual-mode phones that are capable of working on both 2G and 3G systems. The GSM authentication, session key generation, and encryption processes are depicted in Figure 2.

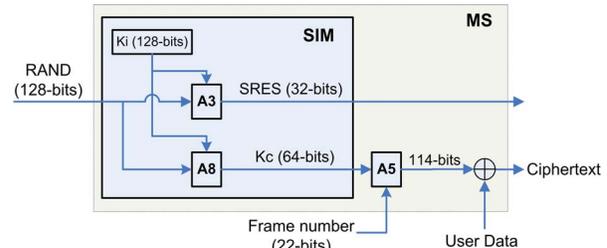

**Figure 2. GSM Authentication, Session key generation, and Ciphering**

The anonymity in the GSM is provided by replacing the use of IMSI with a 32-bit *Temporarily Mobile Subscriber Identity* (TMSI). TMSI is typically handled by the VLR, is valid in a particular *Location Area* (LA), and will be updated at least in every location update procedure. It is also stored on the subscriber's SIM and prevents an eavesdropper to track a particular subscriber.

## 3. Challenges to the GSM Security

The openness of wireless communications makes the communicating parties more vulnerable to the security threats. Although GSM tried to harden the interception by using several techniques such as frequency hopping, the real-time interception of the exchanged information is completely practical [7]. Currently, there are commercial equipments capable of simultaneously intercepting several collocated subscribers [8]. While GSM was intended to be a secure wireless system and considered the user authentication and over-the-air encryption, it is completely vulnerable to several attacks, each of them aiming a part of network. Hereunder, the most important security flaws of the GSM are briefly listed. Several practical scenarios can also be deployed to misuse such vulnerabilities that are neglected for the case of brevity.

**1) Unilateral authentication and vulnerability to the man-in-the-middle attack:** This is the network that authenticates users. The user does not authenticate network so the attacker can use a false BTS with the same mobile network code as the subscriber's legitimate network to impersonate himself and perform a man-in-the-middle attack. The attacker can then perform several scenarios to modify or fabricate the exchanged data. At the designing phase of the GSM protocols, this kind of attack seemed impractical due to costly required equipments. Currently, this kind of attack is completely applicable due to the decreased costs.

**2) Flaws in implementation of A3/A8 algorithms:** Although the GSM architecture allows operator to choose any algorithm for A3 and A8, many operators used COMP128 (or COMP128-1) that was secretly developed by the GSM association. The structure of COMP128 was finally discovered by reverse engineering and some revealed documentations, and many security flaws were subsequently discovered. In addition to the fact that COMP128 makes revealing Ki possible especially when specific challenges are introduced, it deliberately sets ten rightmost bits of Kc equal to zero that makes the deployed cryptographic algorithms 1024 times weaker and more vulnerable, due to the decreased keyspace. Some GSM network operators tried another new algorithm for the A3/A8, called COMP128-2. COMP128-2 was also secretly designed and inherited the problem of decreased keyspace. Despite of such important problem, no other problems are reported so far. However, we can prospect for new discovered vulnerabilities in the future as it is secretly designed. An improved version of COMP128-2, called COM128-3, is also proposed that generates 64 bits of session key and resolves the problem of decreased keyspace.

**3) SIM card cloning:** Another important challenge is to derive the root key Ki from the subscriber's SIM. In April 1998, the *Smartcard Developer Association* (SDA) and the ISAAC research group could find an important vulnerability in the COMP128 algorithm that helped them to extract Ki in eight hours by sending many challenges to the SIM. Subsequently, some other schemes were proposed that were based on the chosen challenges and were capable of extracting Ki in fewer times. Ultimately, a side-channel attack, called partitioning attack, was proposed by the IBM researchers that makes attacker capable of extracting Ki if he could access the subscriber's SIM just for one minute [9]. The attacker can then clone the SIM and use it for his fraudulent purposes. The COMP128 algorithm needs large lookup tables that would leak some important information via the side channels when it is implemented on a small SIM.

**4) Over-the-air cracking:** It is feasible to misuse the vulnerability of COMP128 for extracting the Ki of the target user without any physical access to the SIM. This can be accomplished by sending several challenges over the air to the SIM and analyzing the responses. However, this approach may take several hours. The attacker can also extract IMSI using an approach that will be explained later. After finding Ki and IMSI of the target subscriber, the attacker can clone the SIM and make and receive calls and other services such as SMS in the name of the victim subscriber. However, the attacker will encounter with a slight problem. The GSM network allows only one SIM to access to the network at any given time so if the attacker and the victim subscriber try to access from different locations, the network will realize existence of duplicated cards and disables the affected account.

**5) Flaws in cryptographic algorithms:** Both A5/1 and A5/2 algorithms were developed in secret. The output of A5/1 is the XOR of three LFSRs. An efficient attack to A5/1 that can be used for a real-time cryptanalysis on a PC includes two kinds of attacks [10]: The former that requires the first two minutes of eavesdropped encrypted conversation is capable of extracting the ciphering key in about one second, while the latter just needs two seconds of encrypted conversation to extract the ciphering key in several minutes. A5/2 is the deliberately weakened variant of A5/1. An efficient attack to A5/2 requires less than one second of encrypted conversation to extract the ciphering key in less than one second on a PC [11].

**6) Short range of protection:** The encryption is only accomplished over the airway path between MS and BTS. There is not any protection over other parts of network and the information is clearly sent over the fixed parts. This is a major exposure for the GSM, especially when the communication between BTS and BSC is performed over the wireless links that have potential vulnerabilities for interception. In some countries, the encryption facility of the air interface is not activated at all. There are also security problems on the GSM backbone. The deployed *Signaling System no.7* (SS7) has also several security vulnerabilities. The messages in the current SS7 system is so that they can be modifies or even fabricated into the global SS7 system in an uncontrolled manner [12]. SS7 incorporates very limited authentication procedures since it was originally designed for the closed telecommunication communities. The interconnection with Internet can also have its potential vulnerabilities. Additional vulnerabilities will be arisen when SS7 systems are interconnected using the Internet. Remote management of the GSM backbone elements that can be conducted by connecting them to the IP networks can also introduce additional vulnerabilities. If the HLR and AuC are physically separated, it can be a new point of vulnerability since the authentication triplets may be obtained from AuC by masquerading as another system entity, e.g. a HLR. Unauthorized accesses to HLR, AuC, and MSC will also cause several problems.

**7) Lack of user visibility:** The ciphering is controlled by the BTS. The user is not alerted when the ciphering mode is deactivated. A false BTS can also deactivate the ciphering mode and force MS to send data in an unencrypted manner.

**8) Leaking the user anonymity:** Whenever a subscriber enters a location area for the first time or when the mapping table between the subscriber's TMSI and IMSI is lost, the network requests the subscriber to clearly declare the IMSI. This can be misused to fail the user's anonymity and can be accomplished by sending an IDENTITY REQUEST command from a false BTS to the MS of the target user to find the corresponding IMSI.

**9) Vulnerability to the DoS attack:** A single attacker is capable of disabling an entire GSM cell via a *Denial of Service* (DoS) attack. The attacker can send the CHANNEL REQUEST message to the BSC for several times but he/she does not complete the protocol and requests another signaling channel. Since the number of signaling channels is limited, this leads to a DoS attack. It is feasible since the call setup protocol performs the resource allocations without adequate authentication. This attack is economical since it does not have any charge for the attacker. It can also be used for many practical situations such as terrorist attacks [13].

**10) Absence of integrity protection:** Although the GSM security architecture considers authentication and confidentiality, there is no provision for any integrity protection of information [2]. Therefore, the recipient cannot verify that a certain message was not tampered with.

**11) Vulnerability to replay attacks:** The attacker can misuse the previously exchanged messages between the subscriber and network in order to perform the replay attacks.

**12) Increased redundancy due to the coding preference:** The *Forward Error Correcting* (FEC) is performed prior to the ciphering so there is a redundancy that increases the security vulnerabilities of deployed cryptographic algorithms.

## 4. Security of Transport Channels

The GSM network has some transport channels: *Short Message Service* (SMS), *Unstructured Supplementary Service Data* (USSD), *Wireless Application Protocol* (WAP), and the voice channel. There are also some newer services such as *Enhanced Messaging Service* (EMS) and *Multimedia Messaging Service* (MMS) that have been added in the GSM upgrades. The security flaws described in the previous section are commonly applicable to all the services and transport channels since they aim all the exchanged data and signaling information. However, in addition to such common flaws, some of GSM transport channels have some extra problems and vulnerabilities. The SMS messaging has some extra security vulnerabilities due to its store-and-forward attribute, and the problem of fake SMS that can be conducted via the Internet. When a user is roaming, the SMS content passes through different networks and perhaps the Internet that exposes it to various vulnerabilities and attacks. Another concern is arisen when an adversary gets access to the phone and reads the previous unprotected messages. The USSD that is a session-oriented technology is also vulnerable to attacks since the messages are not encrypted and secured on the GSM backbone.

The WAP allows ME to connect to the Internet. The WAP Gateway that resides between MS and Web server in the WAP architecture and acts as an interpreter between the Internet protocols (HTTP, SSL/TLS, and UDP/TCP/IP) and the corresponding WAP protocols (WSP/WTP, WTLS, and WDP), introduces an additional security flaw in some implementations that is referred to as the WAP gap. Other concerns are arisen from security problems of the Internet as a huge uncontrolled network that is in contradiction with assumptions of the GSM security architecture in which the core network is assumed as a secure and controlled environment. The web servers may also download and execute malicious applets at the client (ME) so the safety of applets and other downloaded programs is another concern.

## 5. Solutions to the GSM Security Flaws

The GSM specifications have been revolutionized during times. In 2002, several efforts have been done to design new cryptographic algorithms for GSM, ECSD, GPRS, and EGPRS that can be implemented on dual-mode phones. Ultimately, A5/3 for GSM and ECSD/EDGE, GEA3 for GPRS, and f8 for UMTS were proposed, all of them having a similar structure. The security mechanisms of the GPRS are similar to that of the GSM. However, instead of using A5 algorithm, GPRS uses the *GPRS Encryption Algorithm* (GEA) that currently has three versions: GEA1, GEA2, and GEA3. In the GPRS, the end terminal of encryption is moved towards a deeper point in the network, i.e. the SGSN. Although the encryption is performed at the physical layer of the GSM, it is accomplished at the *Logical Link Control* (LLL) layer of the GPRS. The UMTS, in addition to its new offered applications, has scrutinized the GSM security problems and has resolved most of them. The main reason of GSM security problems was due to the fact that its security was provided by obscurity so the UMTS algorithms were openly designed. Consequently, its algorithms are not encountered with serious problems. Although some theoretical attacks are proposed, they are not practically feasible with the current technology. However, there are also some other problems related to the deployed protocols.

Regardless of security improvements in the upper generation networks, it is necessary to provide solutions to improve the security of the currently available 2G systems. Hereunder, some practical solutions are proposed for this purpose.

**1) Using secure algorithms for A3/A8 implementations:** This can thwart the dangerous SIM card cloning attack. This solution is profitable since the network operators can perform such improvement themselves and without any need to the software and hardware manufacturers or the GSM consortium. However, this solution requires providing and distributing new SIM cards and modifying the software of the HLR. Currently, both COMP128-2 and COMP128-3 algorithms thwart the SIM card cloning and over-the-air cracking of Ki. Since COMP128-3 enhances the effective key length of the session key to further 10 bits, it allows the deployed cryptographic algorithm to have its nominal security. Although it is soon to judge on the real security of COMP128-2 and COMP128-3, they have apparent advantages over the traditional COMP128-1 that its SIM cloning apparatus are available at very low prices.

**2) Using secure ciphering algorithms:** Operators can use newer and more secure algorithms such as A5/3 provided that such improvements are allowed by the GSM consortium. The deployed cryptographic algorithms should be implemented on both BTS and mobile phones. Any change to the cryptographic algorithms requires agreement and cooperation of software and hardware manufacturers since they should perform the appropriate changes to their products. Since the cryptographic algorithms should be implemented on the cellular phones, the agreement of mobile phone manufacturers is also required. However, a lonely upgrading of the deployed cryptographic algorithms cannot be so useful. Even though the ciphering algorithms are replaced with the strongest ones, the attacker can simply impersonate the real network and force MS to deactivate the ciphering mode so it is also necessary to modify the authentication protocols.

**3) Securing the backbone traffic:** Encrypting the backbone traffic between the network components can prevent the attacker to eavesdrop or modify the transmitted data. Although this solution may be implemented without the blessings of GSM consortium, the cooperation of hardware manufacturers is still required.

**4) End-to-end Security:** The best, easiest, and most profitable solution is to deploy the end-to-end security or security at the application layer. Most of GSM security vulnerabilities (except SIM cloning and DoS attacks) do not aim ordinary people, and their targets are usually restricted to special groups so it is reasonable and economical that such groups make their communications secure by the end-to-end security. Since the encryption and security establishment is performed at the end-entities, any change to the GSM hardware will not be required. In this way, even if the conversation is eavesdropped by the police or legal organizations, they cannot decrypt the transmitted data without having the true ciphering key, provided that a secure enough cryptographic algorithm is deployed. Therefore, in order to avoid illegal activities, it should be transparent to both GSM operator and service provider. It may also be necessary to find solutions for a legal interception or a key screw scheme. The end-to-end security establishment has a complete flexibility to the deployed algorithms so the appropriate upgrades can be easily accomplished when necessary. However, it may be a subject to export control. Generally, the end-to-end security can be provided in the cellular systems by following one or some of the following approaches:

1) *Exploiting the processing capabilities of mobile phone using the programming languages such as J2ME (Java 2 Mobile Edition)*: Supported by the most recent cellular phones and *Personal Digital Assistants* (PDA) with the improved processing capabilities.

2) *Exploiting the processing capabilities of the SIM using the SIM Application Toolkit (SAT)* [14]: Not supported by all SIM cards; especial SIM cards are required; the processing resources are still limited; and operations may be so time-consuming.

3) *Exploiting the processing capabilities of an additional smart card, e.g. JavaCard*: Not supported

by the usual phones; requires costly dual slot cellular phones.

4) *Exploiting the processing capabilities of a portable PC (laptop) connected to the ME*: suitable for security mechanisms with huge processing and memory requirements, e.g. real-time end-to-end secure voice communications over the GSM voice channel [15].

5) *Exploiting the processing capabilities of a crypto-processor that is embedded in the ME* [16]: It should be accomplished by the mobile manufacturer; cannot be changed or manipulated by the user; and may be a subject to export control.

The first four approaches have an inherent advantage due to their capability of being simply manipulated by the end-entities. However, choosing the most profitable approach regards to some parameters such as required memory and processing resources of the corresponding application. For example, if the voice is to be end-to-end encrypted over the data channel, it can even be implemented by a software application that is installed on an advanced cellular phone. On the other hand, for encryption over the voice channel that is hard to be tracked and so attractive for the terrorist and illegal activities, the fourth approach may be suitable [15]. The end-to-end security can be established by both symmetric and asymmetric encryption. The asymmetric encryption is usually too slow to be used for the real-time applications and may be used for the key establishment of a symmetric encryption algorithm. The public keys are usually jointed with the certificates. The private keys and the certificates can be securely stored on either SIM card, an additional smart card (for the dual-slot phones), or a secure hardware on the phone. There are also some proposals for the *Wireless Public Key Infrastructure* (WPKI).

## 6. Conclusions

In this paper, the security of the GSM network is evaluated, and a complete and brief review of its security problems is presented. It is proved that the GSM network has many inherent security flaws that can be misused for fraudulent purposes or for deceiving users. Some practical solutions to improve the security of currently available 2G networks are also proposed. Some solutions include the security improvement of the infrastructure while the others tend to provide the end-to-end security. It is also deduced that the end-to-end security or the security at the application layer is the best and most profitable solution for the currently available 2G systems.